# Record-High Proximity-Induced Anomalous Hall Effect in $(Bi_xSb_{1-x})_2Te_3$ Thin Film Grown on $CrGeTe_3$ Substrate

Xiong Yao,[†] Bin Gao,[‡] Myung-Geun Han,[⊥] Deepti Jain,[§] Jisoo Moon,[§] Jae Wook Kim,[§] Yimei Zhu,[⊥] Sang-Wook Cheong,[†] and Seongshik Oh[*,†]

[†]Center for Quantum Materials Synthesis and Department of Physics & Astronomy, Rutgers, The State University of New Jersey, Piscataway, New Jersey 08854, United States

[‡]Department of Physics and Astronomy, Rice University, Houston, Texas 77005, United States

[⊥]Condensed Matter Physics and Materials Science, Brookhaven National Lab, Upton, New York 11973, United States

[§]Department of Physics & Astronomy, Rutgers, The State University of New Jersey, Piscataway, New Jersey 08854, United States

*Email: ohsean@physics.rutgers.edu

Phone: +1 (848) 445-8754 (S.O.)

ABSTRACT

Quantum anomalous Hall effect (QAHE) can only be realized at extremely low temperatures in magnetically doped topological insulators (TIs) due to limitations inherent with the doping process. In an effort to boost the quantization temperature of QAHE, magnetic proximity effect in magnetic insulator/TI heterostructures has been extensively investigated. However, the observed anomalous Hall resistance has never been more than several Ohms, presumably owing to the interfacial disorders caused by the structural and chemical mismatch. Here, we show that, by growing $(Bi_xSb_{1-x})_2Te_3$ (BST) thin films on structurally and chemically well-matched, ferromagnetic-insulating $CrGeTe_3$ (CGT) substrates, the proximity-induced anomalous Hall resistance can be enhanced by more than an order of magnitude. This sheds light on the importance of structural and chemical match for magnetic insulator/TI proximity systems.

Keywords: magnetic proximity effect, anomalous Hall effect, $CrGeTe_3$, $(Bi_xSb_{1-x})_2Te_3$ , magnetic topological insulator

The discovery of QAHE,[1-6] which exhibits quantized Hall resistance with dissipationless longitudinal transport in the absence of external magnetic field, is one of the major breakthroughs in condensed matter physics in recent years. However, QAHE can be achieved only in magnetically-doped 3D topological insulators,[2, 3] and the full quantization requires extremely low temperatures. Despite improvements with materials engineering schemes such as Cr/V codoping or magnetic modulation doping,[4, 7, 8] full quantization does not occur at temperatures higher than 500 mK. Achieving QAHE at easily accessible temperatures is one of the most important problems in the current QAHE research.

It has been suspected that (both magnetic and compensation) doping is the main factor responsible for the low quantization temperatures. First of all, the amount of magnetic dopant that can be incorporated into the TI films is limited to a small value in order to keep the system in the topological regime.[9, 10] Moreover, the randomly introduced dopants lead to degraded carrier mobility and morphology.[10, 11] Magnetic inhomogeneity, which is inevitable during the doping process, is also believed to negatively affect the quantization temperature.[12-14]

Given the above limitations of the magnetic doping process, an alternative approach to achieve QAHE is to rely on magnetic proximity coupling. Compared with magnetic doping, the proximity coupling can provide more homogeneous exchange interaction. The proximity induced AHE has been observed in several magnetic insulator/TI hybrid systems such as $Bi_2Se_3$/EuS,[15] $(Bi_xSb_{1-x})_2Te_3$/$Y_3Fe_5O_{12}$ (BST/YIG) ,[16] $(Bi_xSb_{1-x})_2Te_3$/$Tm_3Fe_5O_{12}$ (BST/TIG),[17] $Bi_2Se_3$/YIG,[18-20] and $Bi_2Se_3$/$LaCoO_3$.[21] However, the magnetic insulators engaged in these hybrid systems possess quite different crystal structures and anions from TI thin films. The structural and chemical mismatch at the interface between substrate and TI layer inevitably introduces extra defects,[22] deleterious for the realization of large AHE. Even worse, this mismatch can create a chalcogenide-rich dead layer at the interface, undermining the exchange interaction between the magnetic insulator and the TI layer, as evidenced by the fact that the insertion of a 1 nm-thick $AlO_x$ layer can block the magnetic coupling in the $Bi_2Se_3$/YIG heterostructure.[18] To date, the anomalous Hall resistance achieved in the magnetic insulator/TI hybrid system has been limited to very small values, not more than several Ohms. The poor interface quality caused by the structural and chemical incompatibility is likely to be a main factor.

Here we demonstrate by utilizing CGT as the substrate, epitaxial BST/CGT heterostructure with high quality interface can be grown by molecular beam epitaxy (MBE), leading to more than an order magnitude increase in AHE signal than previous values reported in other magnetic insulator/TI systems.

CGT, a van der Waals material, is a ferromagnetic insulator with reported $T_c$ of 61 K-68 K [23-26] and its ferromagnetism can survive down to monolayer.[24] CGT stands out among ferromagnetic insulators being considered for proximity coupled heterostructures with BST because they share similar crystal structures ($R\bar{3}$ vs $R\bar{3}m$) and the same anion (Te). This structural and chemical match provides significant advantage for CGT as the substrate for proximity effect with BST over other candidates such as EuS, YIG (TIG) or $LaCoO_3$. The bulk CGT single crystal also possesses an out of plane magnetization, which is prerequisite for magnetic proximity coupling with TI thin film. Ji, H. et al. previously tried fabrication of $Bi_2Te_3/CrGeTe_3$ heterostructure by metal organic chemical vapor deposition (MOCVD) technique.[23, 27] However, due to either high bulk carrier density of $Bi_2Te_3$ films or some other factors, the observed AHE signal was much less than an Ohm.

The single crystal substrate CGT was grown by the self-flux method: see Supporting Information part 1 for details. The as-grown crystal exhibits very shinny surface with typical size of 4-6 mm, as shown in Figure 1(a). Figure 1(d) gives the magnetization data measured under magnetic field perpendicular to the ab plane at 5 K, exhibiting soft ferromagnetic feature with a saturation field of 0.24 T. As shown in Figure S3, The $T_c$ of CGT substrate can be roughly determined as 65 K from the minimum of the derivative magnetization $dM/dT$ curve, which is in good agreement with the value reported previously.[23-26] The CGT single crystal can be easily cleaved by a Scotch tape and the cleaved fresh surface shows clean, sharp hexagonal edges, as displayed in Figure 1(b). The as-cleaved CGT single crystal substrate was mounted on an $Al_2O_3$ substrate as a holder, then put into the MBE chamber: more details are included in Supporting Information part 2.

The surface quality of the CGT crystal was then checked by *in situ* reflection high-energy electron diffraction (RHEED). At room temperature, the RHEED pattern of CGT substrate is composed of bright streaky lines with some spotty feature. After annealing to 260 ℃, the spotty feature disappears and the pattern becomes more streaky, as shown in Figure 1(c), indicating the atomic flatness of the CGT surface. After the *in situ* annealing of CGT substrate, we grow the BST thin film at 260 ℃: see Supporting Information part 2. Figures 2(c) and (d) give the RHEED patterns of the BST thin film: two kinds of spacing with the relative ratio of $\sqrt{3}$ appear alternately every 30 degree during sample rotation. This represents the highly ordered in-plane orientation of the pseudo-hexagonal BST thin film. It is worth mentioning that the RHEED patterns of BST/YIG,[16] BST/TIG[17] and $Bi_2Se_3$/$LaCoO_3$[21] heterostructures showed coexistence of the two different spacings at the same angle, indicating randomly orientated domains, which is similar to the case of $Bi_2Se_3$ thin films grown on amorphous $SiO_2$.[28] The atomic force microscopy (AFM) images of $Bi_2Se_3$/$LaCoO_3$ also look similar to that of $Bi_2Se_3$ grown on amorphous $SiO_2$,[21, 28] confirming the randomly oriented domains, which are likely induced by the structural mismatch at the interface between the magnetic insulators and TI. Here the RHEED patterns of BST/CGT sample are quite different from those observed previously, proving the improved sample quality by utilizing CGT as the substrate. We checked the surface morphology by AFM after deposition of BST, as shown in Figure 2(b). Highly ordered large flat terraces in characteristic triangular shape with quintuple-layer (QL) steps of 1 nm can be observed in the surface of the BST thin film. The surface morphology of BST on CGT is much better than previous results on other magnetic insulators, suggesting the superior structural and chemical match. Figure 2(e) shows the high angle annular dark-field scanning tunneling electron microscopy (HAADF-STEM) image for a 6 QL BST thin film grown on CGT substrate, exhibiting sharp, disorder-free interface between BST and CGT. Figure 2(f) shows the corresponding elemental mapping electron dispersive X-ray spectroscopy (EDS) images for Cr-Kα (5.412 keV). The Cr content remains almost constant at CGT region then drops to noise level at the BST and Au capping regions: no obvious Cr interdiffusion can be seen in the BST film considering that the Cr content remains the same noise level in both BST and Au regions.

Figure 3(e) gives the temperature dependence of $R_{xx}$ for CGT and BST/CGT samples. CGT bulk crystal is quite conductive at room temperature, but it becomes highly insulating at low temperature. Accordingly the resistance values of the BST/CGT samples are dominated by CGT at temperatures above ~150 K but by BST below ~100 K: thus, we limit our analysis to temperatures below 100 K in order to avoid the substrate contribution. Here the CGT control sample was prepared in the exactly same way as other substrates (mounted on an $Al_2O_3$ substrate, cured with the adhesive, cut into a similar size, cleaved for a fresh surface and *in-situ* annealed at 260 ℃ for 15 min): see Supporting Information part 2 for the detailed sample preparation method.

In order to probe the AHE signal, we performed the Hall measurement on two BST (6 QL)/CGT samples with different carrier types (one p- and the other n-type, denoted as Sample P and Sample N respectively), as shown in Figure 3(a) and (c). We can clearly see a non-linear feature on the Hall resistance $R_{xy}$. For both samples, the observed non-linear Hall feature saturates at 0.24 T and disappears above 61-68 K, which are, respectively, the saturation field and the $T_c$ of CGT. This implies that the observed non-linear Hall effect is due to AHE, i.e., $R_{AH} \propto M$, where M is the magnetization in BST, induced by CGT. Based on fitting of the linear part, the sheet carrier density is $1.9 \times 10^{12}$ /cm$^2$ for Sample P (x = 0.25) and $1.1 \times 10^{12}$ /cm$^2$ for Sample N (x = 0.3). These two values are the lowest among the magnetic insulator/TI heterostructures [16-18, 21], almost comparable to the best BST films grown on $Al_2O_3$ (0001) substrates [29]. The low carrier density is another sign of the improved interface quality with minimal interfacial and bulk defects [22]. To single out the anomalous Hall component from $R_{xy}$, we subtracted the linear background of ordinary Hall effect, and the corresponding $R_{AH}$ are displayed in Figure 3(b) and (d). The maximal $R_{AH}$ reaches 153 Ω for Sample P and 145 Ω for Sample N, more than one order larger than the $R_{AH}$ values reported in previous works.[16-18, 21] Absence of hysteresis in $R_{AH}$ is due to softness of the ferromagnetism and is consistent with the magnetization data (Figure 1(d)) of the CGT substrate.[24]

The following analysis confirms that the large $R_{AH}$ is induced by magnetic proximity coupling rather than Cr interdiffusion. First, the shape of the $R_{AH}$ signal resembles the magnetization data of CGT in Figure 1(d). Moreover, the saturation field of $R_{AH}$ for BST/CGT samples matches almost exactly that (0.24 T) of the magnetization data (Figure 1(d)) for CGT bulk crystal measured at similar temperatures. Anomalous Hall effect can only derive from conducting magnets [30] and so the observed $R_{AH}$ can only come from the TI layer because the CGT substrate is several orders more insulating than the TI layer at the measurement temperature. If the $R_{AH}$ is caused by Cr interdiffusion which acts as Cr doping, then the saturation field of $R_{AH}$ should vary depending on the Bi doping ratio or the level of Cr interdiffusion. However, for all the BST/CGT samples we made, the saturation field always sticks to the same value of 0.24 T. Secondly, the observed $R_{AH}$ for both Sample P and Sample N gradually declines from 153 Ω (145 Ω) at 6.6 K to 84 Ω (97 Ω) at 50 K, then disappears above the $T_c$ of CGT (61-68 K), as shown in Figure 3(f). Even in more complicated multilayer configurations such as a sandwiched structure or a TI layer capped with a magnetic layer, the Cr diffusion is known to be negligible compared with the effect of proximity coupling.[31] Finally, the sign of the observed AHE remains the same in both p- and n-type samples, eliminating the possibility of Lorentz force induced Hall signal by the stray field of the CGT substrate. It is also notable that a recent work reported that CGT flake imprints its magnetization in the AHE of Pt thin film deposited on top,[32] which is consistent with our observation. All these observations strongly suggest that the observed $R_{AH}$ should originate from the proximity coupling between CGT and BST.

When comparing with the QAHE regime, both $R_{AH}$ and $R_{xx}$ should be considered because QAHE requires not only quantized $R_{AH}$ but also vanishing $R_{xx}$. Accordingly, the Hall angle defined as $\tan\theta_H = R_{AH}/R_{xx}$ can be a better parameter than $R_{AH}$ alone when characterizing AHE because it contains the information for both $R_{AH}$ and $R_{xx}$. In order to get comprehensive understanding of the AHE observed in the BST/CGT sample, we compare the anomalous Hall resistance, Hall angle and 2D carrier density of previous magnetic insulator/TI heterostructures with the current data in Figure 4.[16-18, 21] From Figure 4, we can see that the BST/CGT system is significantly better than previous proximity systems, based on both $R_{AH}$ and $\tan\theta_H$ values. The $R_{AH}$ and $\tan\theta_H$ achieved in Sample P (N) are 153 $\Omega$ (145 $\Omega$) and 0.01286 (0.00715), respectively, more than one order larger than previous proximity-coupled systems. The high quality interface, achieved by combination of structural and chemical match, is likely to be the main factor leading to the enhanced AHE in the BST/CGT system. One limitation of the CGT system is that it lacks magnetic hysteresis, which is a necessary condition for QAHE.

In conclusion, despite lack of magnetic hysteresis, CGT, which shares similar crystal structure and an identical anion with BST, provides significantly better interface for BST than other magnetic insulators, leading to much higher AHE signals than before. This observation manifests the importance of structural and chemical match toward QAHE through proximity coupling. What's more, proximity coupling was found to enhance or tailor the magnetic ordering in magnetic TI thin films.[20, 33, 34] Developing structurally and chemically compatible magnetic substrates could pave a new avenue for boosting the QAHE temperature or exploring novel quantum phenomenon in magnetic insulator/TI hybrid systems.

**Growth method**: See Supporting Information Part 1 and 2.

**Transport measurement:** All transport measurements were performed with the standard van der Pauw geometry in a closed-cycle cryostat of magnetic fields up to 0.6 T and temperatures down to 6.0 K. Electrical electrodes were made by manually pressing four indium wires on the corners of each sample. All the samples were carefully cut into square shape to minimize the deviation from van der Pauw geometry. Raw data of $R_{xx}$ and $R_{xy}$ were properly symmetrized and antisymmetrized respectively. 2D carrier density was extracted from $n_{2D} = (e \frac{dR_{xy}}{dB})^{-1}$, where e is the electronic charge and $\frac{dR_{xy}}{dB}$ is the slope of the Hall resistance vs magnetic field B, measured at the origin.

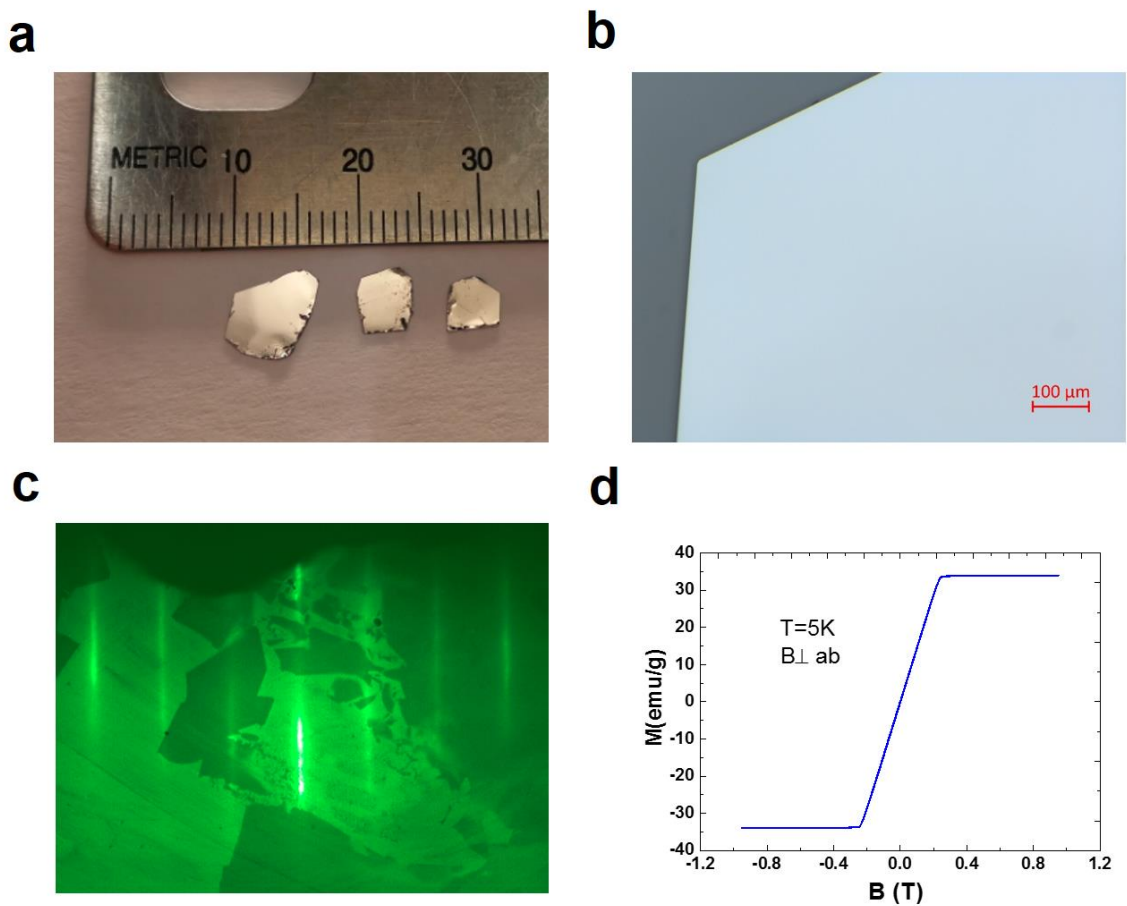

Figure 1: Properties of the single crystal substrate CrGeTe₃. (a) Image of some representative as-grown CGT single crystal samples: typical size of 4-6 mm. (b) Optical microscope image of cleaved fresh surface of a CGT single crystal, showing sharp hexagonal edges indicating the high quality of the crystal. (c) *In situ* RHEED pattern of the cleaved CGT substrate surface, after annealed at 260 ℃ for 15 minutes. The complete RHEED patterns consist of two spacings with a relative ratio of $\sqrt{3}$, repeating every 30 degrees of the sample rotation: see Supporting Information part 2. (d) Magnetic field dependence of magnetization for a CGT substrate measured at 5 K, with the magnetic field perpendicular to the ab plane.

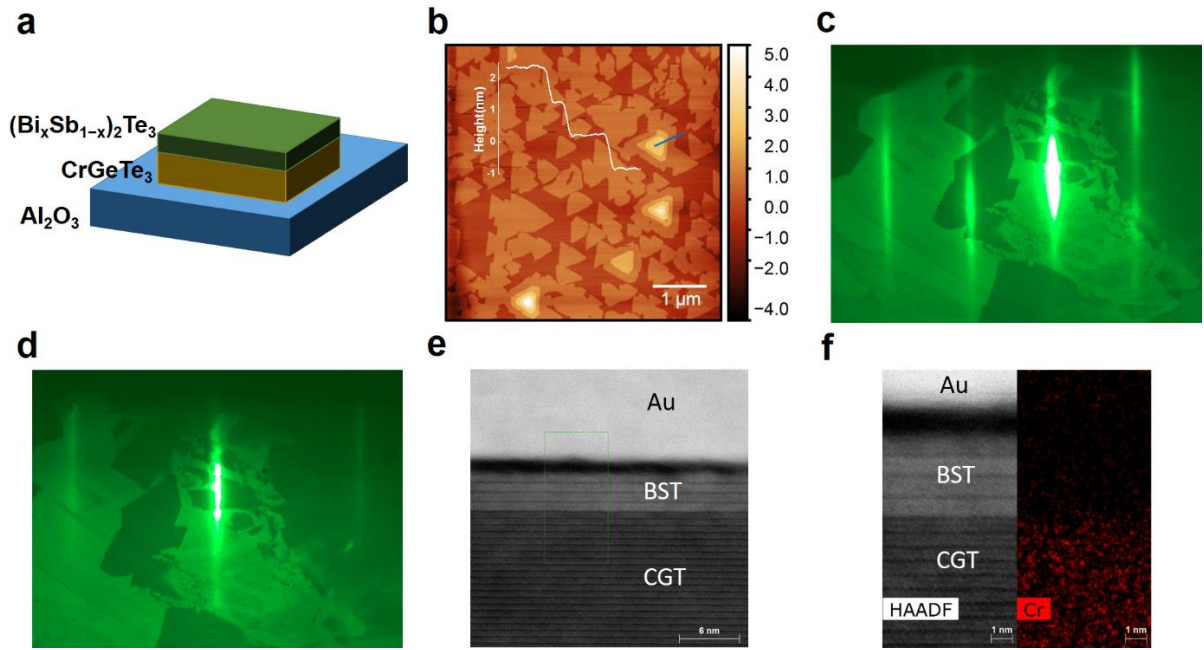

Figure 2: High quality epitaxial BST thin film grown on CGT substrate by MBE. (a) Schematic illustration of the MBE growth for BST/CGT heterostructure: the typical size of CGT substrate is 4-6 mm, and the $Al_2O_3$ substrate only acts as a holder because it fits well with our MBE sample plate. (b) AFM image (5 μm × 5 μm) of a 12 QL BST thin film grown on CGT substrate: the AFM measurement is by tapping mode. Inset shows the height profile of the terrace along the line drawn on the image. Each step is roughly 1 nm, corresponding to the height of 1 QL BST. Note that the films surface is dominated by single-step large terraces. (c-d) Two sets of RHEED spacings appear every 30 degrees when rotating the BST film: the relative ratio of the two spacings is $\sqrt{3}$, corresponding to the hexagonal symmetry. (e) HAADF-STEM image for 6QL BST thin film grown on CGT substrate. The top few BST layers were damaged during Au sputtering for STEM sample preparation, as can been seen from the gap between BST film and Au capping layer. (f)

The left panel shows enlarged view of the green box in (e). The right panel shows the elemental mapping EDS images for Cr (Kα = 5.412 keV).

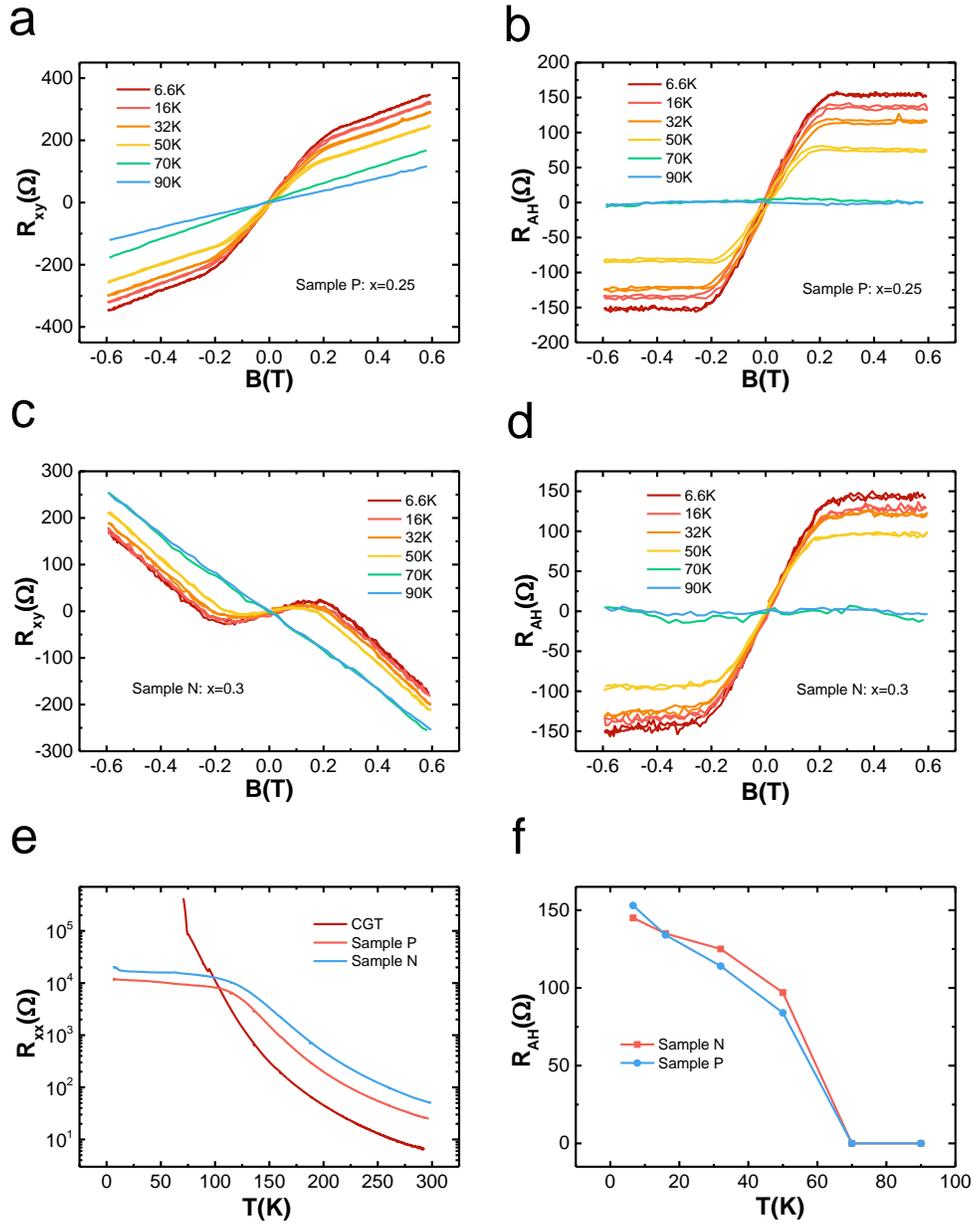

Figure 3: Transport properties of two BST(6 QL)/CGT samples: Sample P and Sample N. All transport measurements were performed with the standard van der Pauw geometry. All the samples

presented here are without a capping layer. (a) and (c) Hall resistance of BST(6 QL)/CGT Sample

P (x = 0.25) and Sample N (x = 0.3) from 6.6 K to 90 K. (b) and (d) Anomalous Hall resistance

for Sample P and Sample N from 6.6 K to 90 K after subtracting the linear background of ordinary

Hall resistance. (e) The longitudinal sheet resistance $R_{xx}$ for a CGT substrate, Sample P, and

Sample N. The $R_{xx}$ values at the high temperature range are dominated by the CGT substrates but differ from one another due to differences in uncontrollable details (such as thickness) between different CGT substrates. (f) Summary of the temperature dependent anomalous Hall resistance for both Sample P and Sample N.

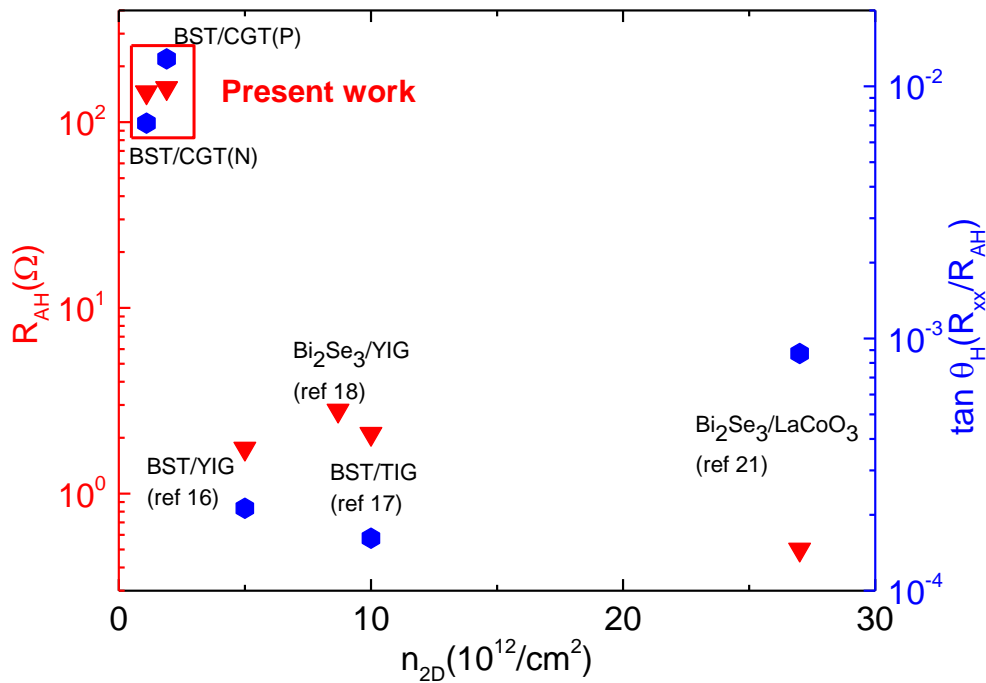

Figure 4: Comparison of present work with other magnetic insulator/TI proximity systems. The

anomalous Hall resistance, Hall angle and 2D carrier density for various magnetic insulator/TI

proximity systems. All data were extracted from the best sample in Ref 16, 17, 18 and 21.

ASSOCIATED CONTENT

**Supporting Information**.

Growth methods for CGT single crystal substrate and BST thin film, temperature dependence of magnetization for CGT substrate, extended AFM results for BST thin films at larger scales, transport results for a 5 QL BST/CGT (x = 0.25) sample, longitudinal sheet resistance for all BST/CGT samples, experimental details about STEM measurements. This material is available free of charge via the Internet at http://pubs.acs.org.

AUTHOR INFORMATION

Corresponding Author

*E-mail: ohsean@physics.rutgers.edu

Author Contributions

X.Y., S.C. and S.O. conceived the experiments. X.Y., B.G. and S.C. grew the CGT substrates. X.Y., D.J. and J.M. grew the thin films. X.Y. performed the transport and AFM measurements, and analyzed the data with S.O. J.K. performed the magnetization measurements of CGT. M.H. and Y.Z. performed the TEM measurements. X.Y. and S.O. wrote the manuscript with contributions from all authors.

Notes

The authors declare no competing financial interest.

ACKNOWLEDGMENT

This work is supported by the center for Quantum Materials Synthesis (cQMS), funded by the Gordon and Betty Moore Foundation's EPiQS initiative through grant GBMF6402, and by Rutgers University. It is additionally supported by Gordon and Betty Moore Foundation's EPiQS initiative through grant GBMF4418 for SO. The work at Brookhaven National Laboratory is supported by the U.S. DOE Basic Energy Sciences, Materials Sciences and Engineering Division under Contract No. DESC0012704. TEM sample preparation is carried out at Center for Functional Nanomaterials, Brookhaven National Laboratory.

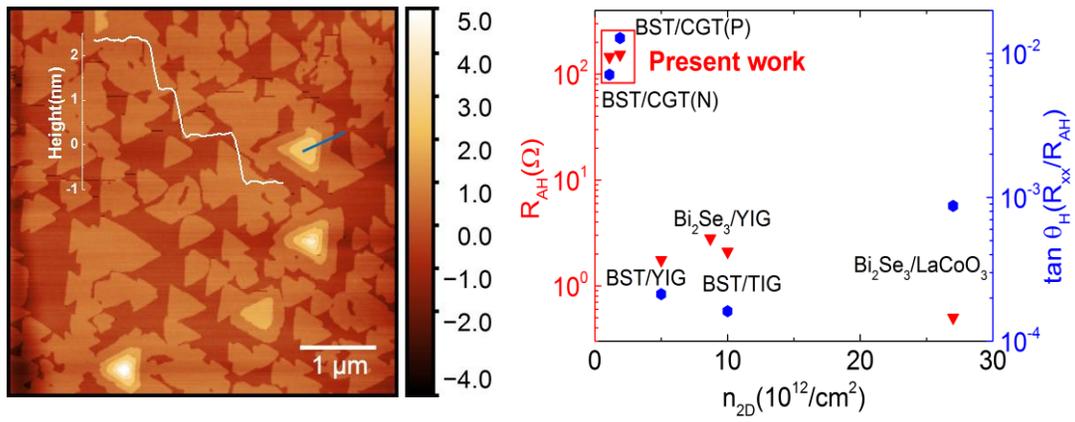

For TOC only

# Supporting Information:
# Record-High Proximity-Induced Anomalous Hall Effect in $(Bi_xSb_{1-x})_2Te_3$ Thin Film Grown on $CrGeTe_3$ Substrate

Xiong Yao,[†] Bin Gao,[‡] Myung-Geun Han,[⊥] Deepti Jain,[§] Jisoo Moon,[§] Jae Wook Kim,[§] Yimei Zhu,[⊥] Sang-Wook Cheong,[†] and Seongshik Oh[*,†]

[†]Center for Quantum Materials Synthesis and Department of Physics & Astronomy, Rutgers, The State University of New Jersey, Piscataway, New Jersey 08854, United States

[‡]Department of Physics and Astronomy, Rice University, Houston, Texas 77005, United States

[⊥]Condensed Matter Physics and Materials Science, Brookhaven National Lab, Upton, New York 11973, United States

[§]Department of Physics & Astronomy, Rutgers, The State University of New Jersey, Piscataway, New Jersey 08854, United States

*Email: ohsean@physics.rutgers.edu

Phone: +1 (848) 445-8754 (S.O.)

# 1. Growth method for CrGeTe₃(CGT) single crystal

High quality CGT single crystals were grown with the self-flux method. Starting materials of Cr (99.995%, Alfa Aesar), Ge (99.999%, Alfa Aesar) and Te (99.999%, Alfa Aesar) with the molar ratio of 1:3:18 were mixed together and then sealed into an evacuated quartz tube. The top part of the quartz tube was filled with some quartz wool as the medium for centrifugation, as illustrated in Figure S1. The quartz tube was put into a box furnace, heated up to 700 ℃, held for 10 hours, and slowly cooled down to 480 ℃ over 3 days. Then the quartz tube was taken out at 480 ℃ from the furnace, flipped over, gently dropped into a centrifugal tube, and then run for 30 seconds.

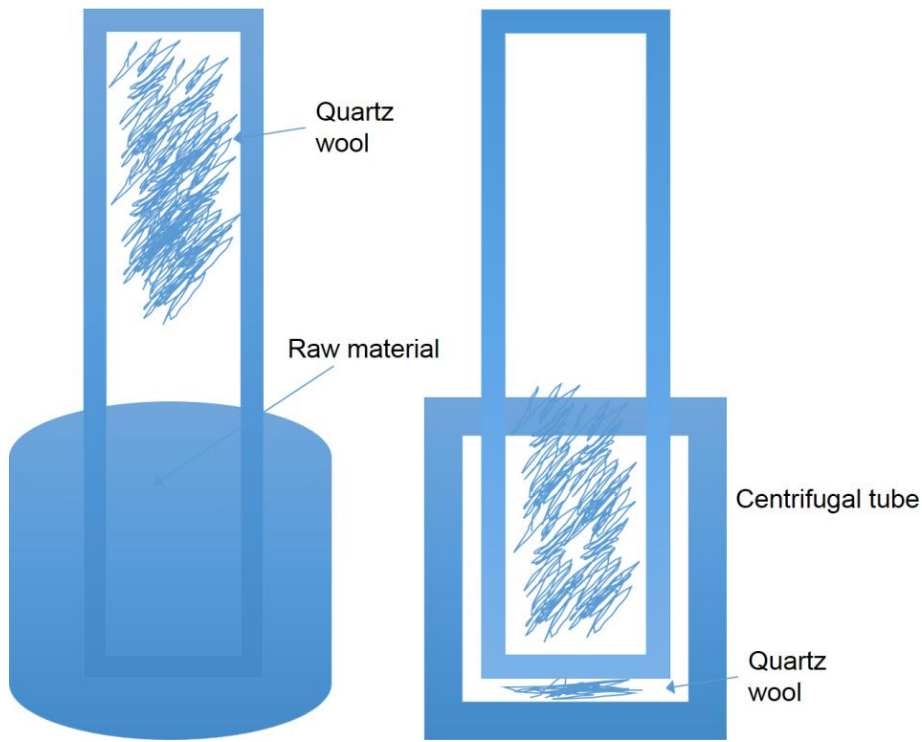

Figure S1: Experimental setup for CrGeTe₃ single crystal growth

## 2. Growth method for $(Bi_xSb_{1-x})_2Te_3$ thin film

The CGT substrate was mounted on a 1 cm × 1 cm $Al_2O_3$ substrate by PELCO® High Performance Ceramic Adhesive (Ted Pella): $Al_2O_3$ only acts as a substrate holder because it fits our standard MBE sample plate. Then the $Al_2O_3$ with CGT on it was put onto a hot plate, heated up to 100 ℃ and stayed for 2 hours to cure the ceramic adhesive. After curing, the bonding between $Al_2O_3$ and CGT becomes very strong so that we can cleave a fresh surface of the CGT. Then we put the substrate into a custom-designed SVTA MOSV-2 MBE system right away. We found 100 ℃ is safe enough to keep the CGT substrate from degradation, and even after the cleaved surface is exposed in the air for a few seconds, oxidation of the surface is almost negligible as indicated by the sharp RHEED pattern of the CGT substrate.

The CGT substrate was *in situ* annealed to 260 ℃ for 15 minutes to improve the surface flatness. *In situ* RHEED images were taken during the annealing, as shown in Figure S2. At room temperature, the RHEED patterns show two kinds of spacing with a relative ratio of $\sqrt{3}$ when rotating the substrate for every 30 degrees. The RHEED patterns are a little spotty at room temperature, indicating the presence of 3D-like features. After annealing, the RHEED patterns become streaky with Kikuchi lines, implying improved surface flatness. For $(Bi_xSb_{1-x})_2Te_3$ thin film deposition, high-purity Bi (99.999%, Alfa Aesar), Sb (99.9999%, Alfa Aesar), and Te (99.9999%, Alfa Aesar) were evaporated from effusion cells. The temperature of CGT substrate was kept at 260 ℃ during the growth. All the fluxes were calibrated by *in situ* quartz crystal microbalance (QCM) and *ex situ* Rutherford backscattering spectroscopy (RBS).

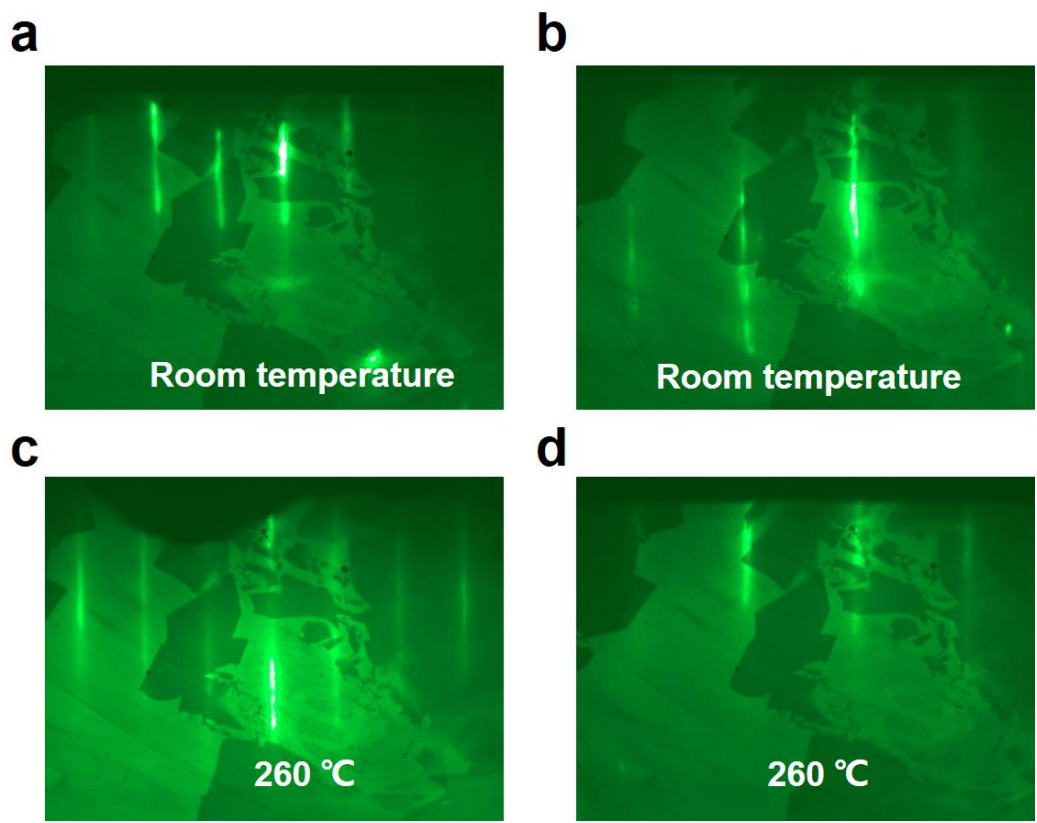

Figure S2: RHEED patterns of a CGT substrate during the annealing process. (a) and (b) Two sets of RHEED patterns showing two different spacings at room temperature. (c) and (d) Improved RHEED patterns after annealing at 260 ℃.

**3. Temperature dependence of magnetization for CGT substrate**

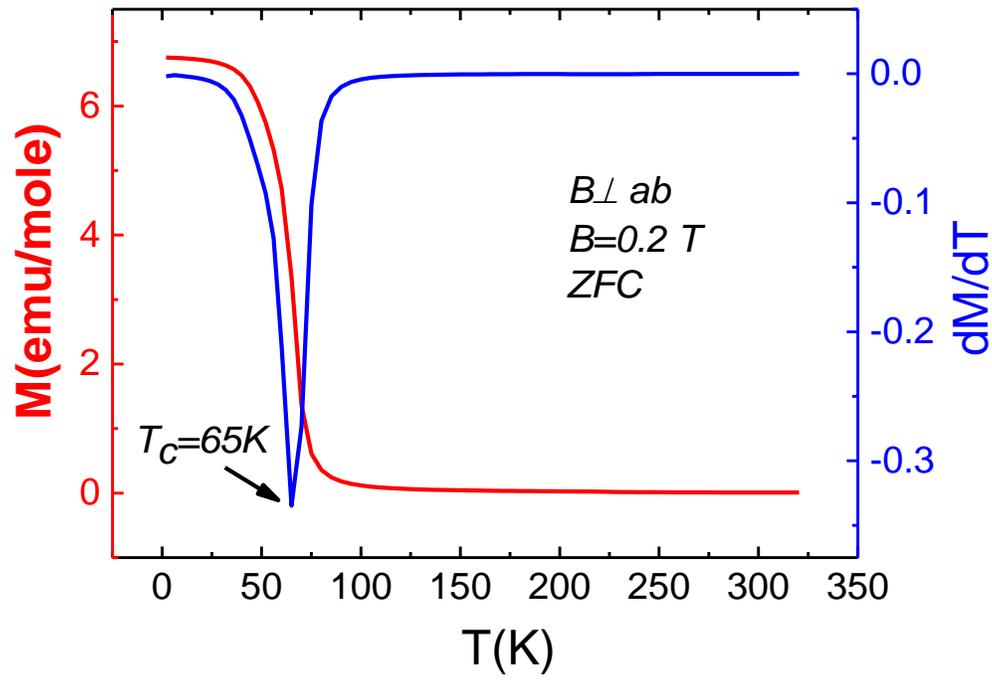

Figure S3: Temperature dependence of magnetization for CGT substrate measured at magnetic field B = 0.2 T, perpendicular to the ab plane. The derivative (dM/dT) vs T is also shown together.

**4. Extended AFM results**

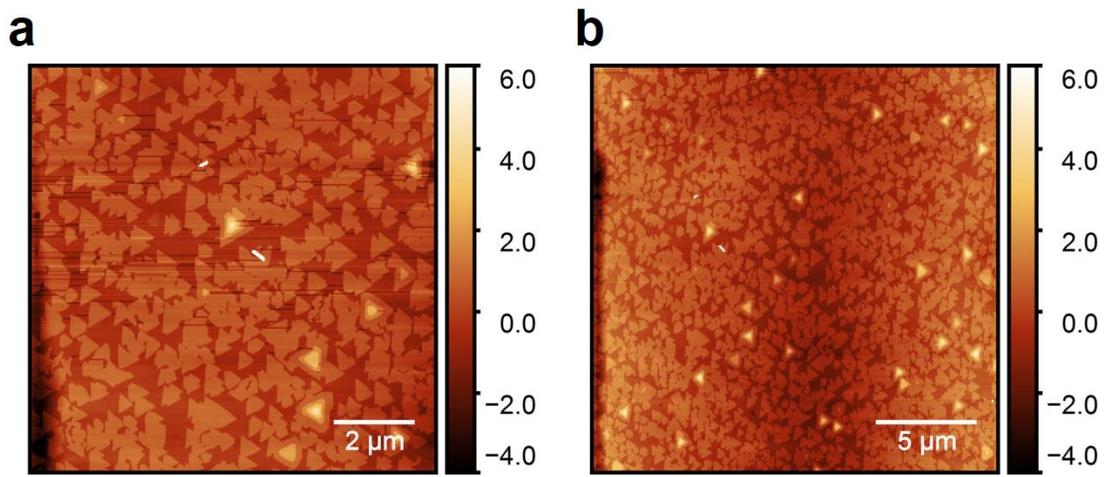

Figure S4: AFM results of a BST(12 QL)/CGT sample at (a) 10 μm × 10 μm and (b) 20 μm × 20 μm scale. Both images indicate homogenous distribution of ordered terraces. All the AFM measurements are by tapping mode.

## 5. Transport results for a 5 QL BST/CGT(x = 0.25) sample

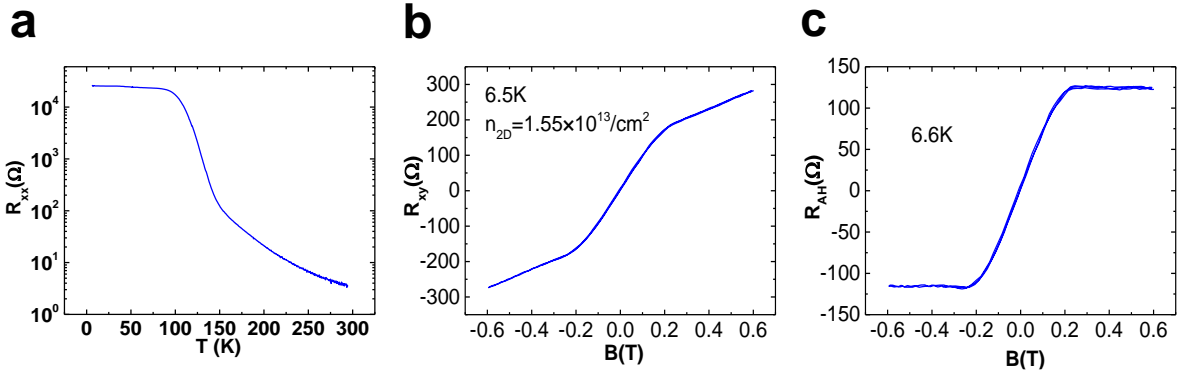

Figure S5: (a) Temperature dependence of longitudinal sheet resistance. (b) Magnetic field dependence of Hall resistance at 6.5 K before removing the linear background. Inset shows the 2D carrier density extracted from the fitting of the linear background. (c) Anomalous Hall resistance after the linear background is removed.

## 6. Longitudinal sheet resistance for all BST/CGT samples

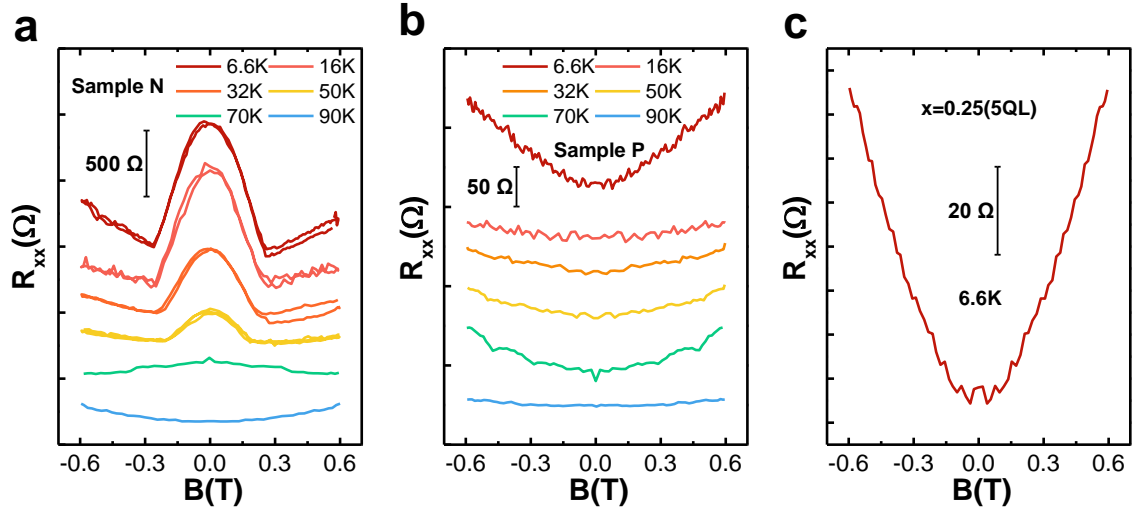

Figure S6: Magnetic field dependence of longitudinal sheet resistance at different temperatures for (a) Sample N (x = 0.30, 6 QL), (b) Sample P (x = 0.25, 6 QL) and (c) x = 0.25 (5 QL): p-type.

Above the coercive field (or above the ferromagnetic $T_C$), all the samples exhibit the standard positive magnetoresistance, which is universally observed for all topological insulators. However, below the coercive field, the magneto-resistance becomes negative for the n-type sample whereas it remains unchanged for the p-type samples. Finding the exact origin behind this anomaly is beyond the scope of the current work, but it seems to be related to the fact that the polarity of anomalous vs ordinary Hall effect is opposite for the n-type samples, whereas it is the same for the p-type samples. Further studies will be needed to fully resolve this question.

**7. Scanning transmission electron microscopy and energy dispersive X-ray spectroscopy**

TEM samples were prepared by focused ion beam lift-out technique with 5 keV Ga+ ions for final milling. An FEI Talos F200X with a four-quadrant 0.9-sr energy dispersive X-ray spectrometer is used for high-angle annular dark-field (HAADF) scanning transmission electron microscopy (STEM) imaging and elemental mapping. The range of collection angles for HAADF STEM imaging was about 70 ~ 280 mrad. For Cr elemental mapping, Cr-Kα (5.412 keV) was used.